# Strain engineering of epitaxial oxide heterostructures beyond substrate limitations

Xiong Deng[1,10], Chao Chen[1,10], Deyang Chen[1,2,11,*], Xiangbin Cai[3], Xiaozhe Yin[1], Chao Xu[4], Fei Sun[1], Caiwen Li[1], Yan Li[5], Han Xu[6], Mao Ye[7], Guo Tian[1], Zhen Fan[1], Zhipeng Hou[1], Minghui Qin[1], Yu Chen[5], Zhenlin Luo[6], Xubing Lu[1], Guofu Zhou[2,8], Lang Chen[7], Ning Wang[3], Ye Zhu[4], Xingsen Gao[1], Jun-Ming Liu[1,9]

[1]Institute for Advanced Materials, South China Academy of Advanced Optoelectronics, South China Normal University, Guangzhou 510006, China

[2]Guangdong Provincial Key Laboratory of Optical Information Materials and Technology, South China Academy of Advanced Optoelectronics, South China Normal University, Guangzhou 510006, China

[3]Department of Physics and Center for Quantum Materials, The Hong Kong University of Science and Technology, Clear Water Bay, Kowloon, Hong Kong, China

[4]Department of Applied Physics, The Hong Kong Polytechnic University, Hung Hom, Kowloon, Hong Kong, China

[5]Institute of High Energy Physics, Chinese Academy of Sciences, Beijing 100049, China

[6]National Synchrotron Radiation Laboratory, University of Science and Technology of China, Hefei, Anhui 230026, China

[7]Department of Physics, Southern University of Science and Technology, Nanshan District, Shenzhen, Guangdong 518055, China

[8]National Center for International Research on Green Optoelectronics, South China Normal University, Guangzhou 510006, China

[9]Laboratory of Solid State Microstructures and Innovation Center of Advanced Microstructures, Nanjing University, Nanjing 210093, China

[10]These authors contributed equally

[11]Lead Contact

*Correspondence:
deyangchen@m.scnu.edu.cn (D.C.)

# SUMMARY

The limitation of commercially available single-crystal substrates and the lack of continuous strain tunability preclude the ability to take full advantage of strain engineering for further exploring novel properties and exhaustively studying fundamental physics in complex oxides. Here we report an approach for imposing continuously tunable, large epitaxial strain in oxide heterostructures beyond substrate limitations by inserting an interface layer through tailoring its gradual strain relaxation. Taking $BiFeO_3$ as a model system, we demonstrate that the introduction of an ultrathin interface layer allows the creation of a desired strain that can induce phase transition and stabilize a new metastable super-tetragonal phase as well as morphotropic phase boundaries overcoming substrate limitations. Furthermore, continuously tunable strain from tension to compression can be generated by precisely adjusting the thickness of the interface layer, leading to the first achievement of continuous orthorhombic (O) - rhombohedral-like (R) – tetragonal-like (T) phase transition in $BiFeO_3$ on a single substrate. This proposed route could be extended to other oxide heterostructures, providing a platform for creating exotic phases and emergent phenomena.

**Progress and Potential**

Epitaxial strain, imparted by an underlying substrate, is a powerful pathway to drive phase transitions and dramatically alter properties in complex oxides, enabling the emergence of new ground states and the enhancement of ferroelectricity, piezoelectricity, superconductivity and ferromagnetism. To realize these emergent phenomena, the availability of appropriate single-crystal substrates for the growth of high-quality epitaxial oxide films with a desired strain state cannot be overemphasized. However, the limitation of commercially available single-crystal substrates and the lack of continuous strain tunability result in stringent restrictions for the further discovery of novel properties and the exhaustive study of fundamental physics. Here, we propose a strategy for imposing continuously tunable, large biaxial strain beyond substrate limitations *via* inserting an interface layer, enabling the achievement of continuous O-R-T phase transition in $BiFeO_3$ thin films on a single substrate and the integration of relevant morphotropic phase boundary on different substrates. This work provides a framework for the strain engineering of complex oxides.

# INTRODUCTION

Strong coupling and complex interplay between strain and spin, charge, orbital and lattice degrees of freedom provide a fertile new ground for creating exotic phases and realizing novel functionalities in complex oxide thin films and heterostructures[1-8]. This has enabled epitaxial strain as a powerful tool for the creation of new ground states (associated with phase transitions)[9-13], the improvement of catalytic activity[14-16], and the manipulation of electric and magnetic properties[17-27] including greatly enhanced superconducting,[28,29] ferroelectric[13,30] and ferromagnetic[31-33] transition temperatures. To realize these emergent phenomena, the availability of appropriate single-crystal substrates for the growth of high-quality epitaxial oxide films with a desired strain state cannot be overemphasized[4,34]. However, both the achievement and the tunability of strain states are hindered owing to the lack of commercially available substrates, thereby resulting in stringent restrictions for the further discovery of novel properties and the exhaustive study of fundamental physics in oxide thin films and heterostructures[4,34-36]. Despite the recent efforts to obtain continuous strain tunability by optical pulses,[37] ion implantation,[38-40] external stress using flexible substrates,[41] piezoelectric-based apparatus,[42] strain-releasing buffer layer[43-45] and thermal expansion mismatch[46], these approaches remain limited either by tuning uniaxial strain only, or by imparting moderate strain values (< 1%), precluding the ability to take full advantage of strain engineering. Therefore, two broad questions remain: (i) are there pathways to obtain the desired strain states overcoming substrate limitations, and (ii) is it possible to achieve continuous strain tunability? Here we report a route for imposing continuously tunable, large biaxial strain in oxide heterostructures beyond substrate limitations *via* the introduction of an interface layer.

BiFeO$_3$ (BFO) is a room temperature multiferroic material that possesses coupled ferroelectric and antiferromagnetic orders[47-49]. The parent ground state of this material is rhombohedral structure (with lattice parameter ~ 3.96 Å)[47]. Epitaxial strain has been exploited to drive rhombohedral-like (R) to tetragonal-like (T) phase transition[3,9,50,51], leading to a wealth of striking observations, such as giant spontaneous polarization of T phase[52], strong piezoelectricity[53,54] and magnetism near R/ T morphotropic phase boundary[55], and the enhancement of anisotropic photocurrent in mixed phase areas[56] (Please note that we denote "R" as rhombohedral-like phase, "T" for the highly distorted tetragonal-like M$_C$ phase, and "T$_R$" for the real tetragonal phase in this paper). However, this strain-induced R-T phase transition and the stabilization of the metastable T phase BFO (T-BFO, with in-plane lattice parameter a ~3.66 Å, out-of-plane lattice parameter c ~4.65 Å and large c/a ratio~1.27), can only be achieved on substrates that provide large compressive strains exceeding -4.3% at room temperature[9,57], whereas films on substrates with epitaxial strain of -4.3% ~ +1% is rhombohedral-like BFO (R-BFO)[9], and films on substrates with tensile strain larger than +1% is orthorhombic phase BFO (O-BFO)[58]. On the other hand, although both R-T and R-O phase transitions have been experimentally induced by large compressive[9,50-52] and tensile epitaxial strain[58], respectively, the strain-driven continuous O-R-T phase transition in BFO remains elusive, which would enable the possibility of designing multiple morphotropic phase boundaries.

In this study, taking BiFeO$_3$ (BFO) as a model system, we demonstrate that the insertion of an ultrathin interface layer (as thin as 7 unit-cells) allows the creation of a desired strain overcoming substrate limitations. Using a combination of X-ray diffraction (XRD) associate with reciprocal space mapping (RSM), scanning probe microscopy (SPM) and transmission electron microscopy (TEM), we find that the use of this approach not only enables the growth of high-quality BFO epitaxial thin films, but also renders the ability to generate a desired strain that can induce phase transition and stabilize the T phase as well as morphotropic phase boundary, no matter on compressive or tensile strain substrates. Moreover, continuously tunable strain cutting across tension to compression can be imparted by accurately adjusting the thickness of an interface layer, enabling the realization of continuous O-R-T phase transition in BFO on a single substrate for the first time.

## RESULTS AND DISCUSSION

**Imposing epitaxial strain beyond substrate limitations**

Based on previous first-principle calculations and phase field simulations[9,56,57], the room temperature strain phase diagram of BiFeO$_3$ is sketched as presented in Figure 1A, showing that R phase forms in moderate epitaxial strains (such as on (001) (LaAlO$_3$)$_{0.29}$–(SrAl$_{1/2}$Ta$_{1/2}$O$_3$)$_{0.71}$ (LSAT, a = 3.868 Å) and (001) SrTiO$_3$ (STO, a = 3.905 Å) substrates) and transforms to O phase under large tensive strain (such as on (110)$_O$ NbScO$_3$ (NSO, a = 4.01 Å) substrate, "O" refers to the orthorhombic structure of NSO and the demonstration of the orthorhombic symmetry will be discussed later), while T-BFO can only be obtained under large compressive strain state (such as on (001) LaAlO$_3$ (LAO, a = 3.79 Å) substrate). By introducing an interface layer Ca$_{0.96}$Ce$_{0.04}$MnO$_3$ (CCMO, a = 3.77 Å)[59], we expect to overcome substrate limitations to impart epitaxial strain for the formation of the metastable T-BFO not only on LAO substrate, but on LSAT and STO substrates and even on NSO (Figure. 1B).

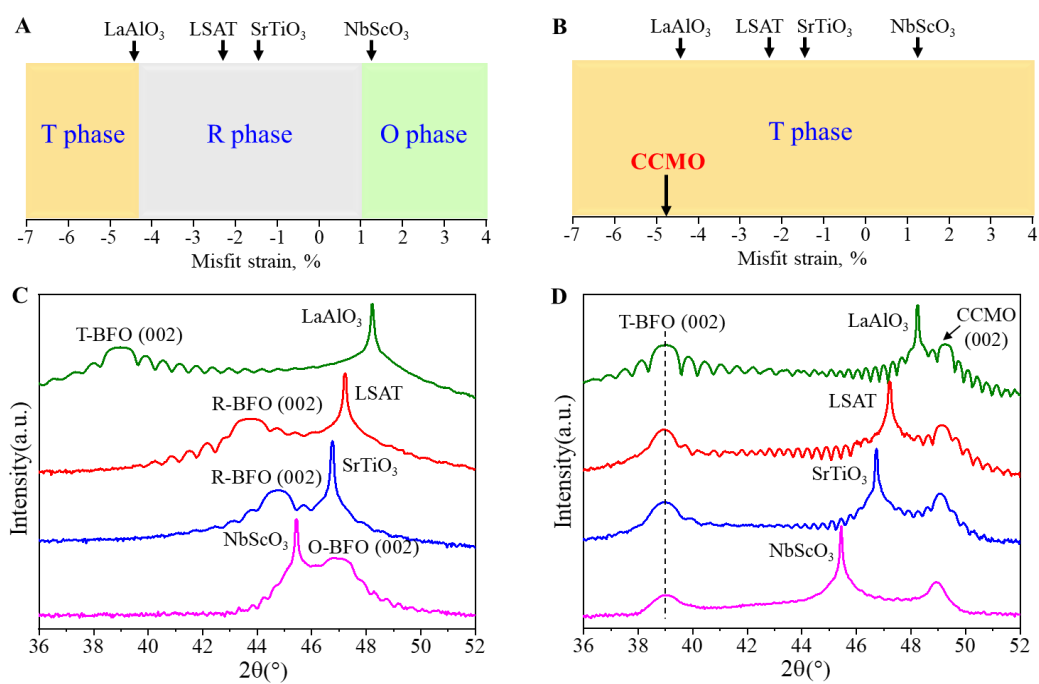

**Figure 1. Strain Engineering within and beyond Substrate Limitations**
(A) Room temperature epitaxial strain phase diagram of BFO shows strain engineered structural changes within substrate limitations.
(B) Expected sketch of the formation of metastable T phase beyond substrate limitations through inserting an interface CCMO layer.
(C) Typical x-ray θ–2θ scans of 14 nm-thick BFO films grown on bare substrates confirm the formation of O, R, T phases induced by different strain state.
(D) XRD data of BFO thin films show the achievement of the metastable T phase on various substrates with the insertion of an interface CCMO layer, revealing the ability to impart epitaxial strain beyond substrate limitations.

Using pulsed laser deposition (PLD), we grew a series of epitaxial BFO thin films and BFO/CCMO heterostructures on (001) LAO, (001) LSAT, (001) STO and (110)$_O$ NSO substrates (Please note that the subscript "O" refers to the orthorhombic structure and all substrates used in this study are pseudocubic (001) oriented single crystals). Typical x-ray θ–2θ scans of 14 nm-thick BFO films grown on these bare substrates illustrate the epitaxial growth of BFO and the creation of T phase on LAO, R phase on LSAT and STO, and O phase on NSO (Figure 1C). These results are consistent with previous theoretical and experimental studies[9,49,50,56,57]. Interestingly, the insertion of a 27 nm-thick CCMO layer (the atomically flat morphology of CCMO is shown in Figure S1) between substrates and BFO films allows us to stabilize the metastable T phase on all these substrates from compression to tension (Figure 1D), demonstrating the ability to impose epitaxial strain beyond substrate limitations. Here we infer that the strain relaxation of CCMO leads to the creation of large compressive strain that enables the stability of T phase.

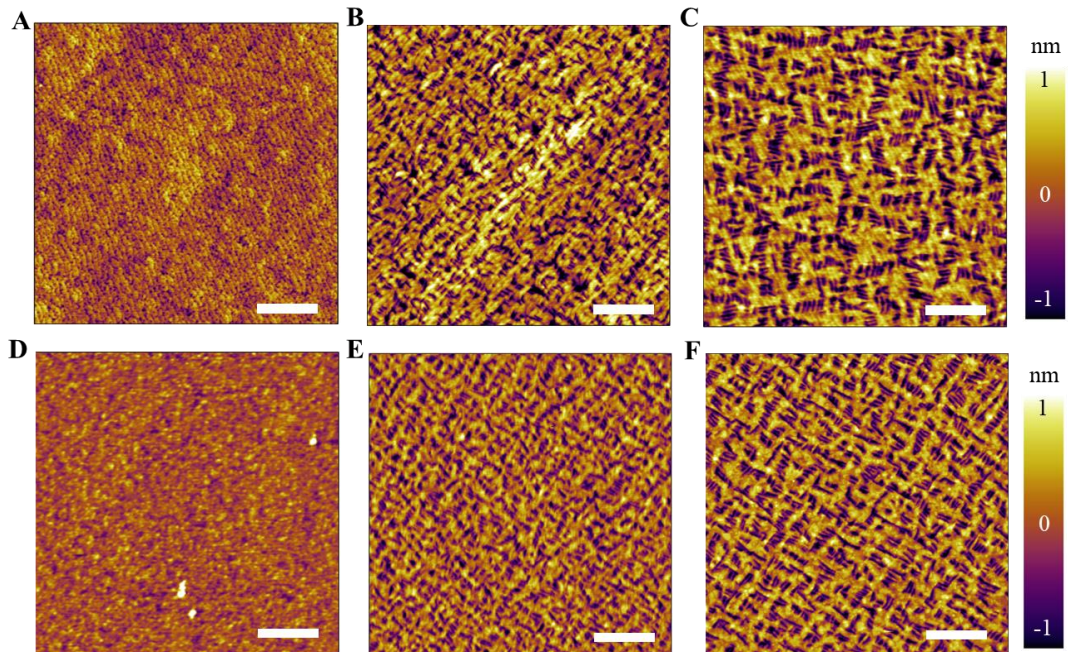

**Figure 2. Thickness Dependent T-R Phase Transition**
AFM images of 7 nm (A, D), 25 nm (B, E) and 50 nm (C, F) thick BFO with 27 nm-thick CCMO buffer layer on $SrTiO_3$ (A-C) and $NbScO_3$ (D-F) substrates, respectively. The scale bars are 1μm.

**Thickness-dependent and electrical control of phase transition**

Having determined that the BFO/CCMO heterostructures grown on LSAT, STO and NSO substrates can induce T-BFO that is analogous to previously observed T phase on LAO[9,53], we ask whether it could also evolve to R/T mixed phase and design morphotropic phase boundary by increasing the film thickness. Atomic Force Microscope (AFM) measurement is performed on a series of BFO/CCMO heterostructures with 7, 25 and 50 nm-thick BFO and 27 nm-thick CCMO on LSAT, STO and NSO substrates. Thickness dependence evolution of surface morphologies (Figure 2) reveal high quality epitaxial growth of BFO films, showing atomically flat morphologies (root mean square (RMS) roughness < 0.3 nm). Furthermore, the one unit-cell in height terraces (Figure 2A) indicate the formation of pure T phase in 7 nm-thick BFO on STO, whereas bright contrast matrix and dark contrast stripes in 25 nm-thick film (Figure 2B) suggest the emergence of the R phase that coexists with the T phase due to relaxation of the epitaxial strain with increased thickness. Further strain relaxation in thicker film (50 nm) results in further increase of fraction of R phase (Figure 2C). The distinct thickness-dependent phase evolution is observed on NSO (Figure 2D-F) and LSAT (Figure S2) substrates as well.

It is worth to mention that here we first time report the achievement of R and T mixed phases and R/T morphotropic phase boundary on LSAT, STO and NSO substrates. These observations are additional evidences for the epitaxial strain induced T phase beyond substrate limitations, in agreement with our XRD data (Figure 1D). The ability to maintain a high compressive strain state on various substrates can not only drive an increased ferroelectric transition temperature in ferroelectric perovskite that is needed for new developments such as in situ polarization monitoring[60,61], but also may promote the integration of perovskite ferroelectrics on silicon[62,63].

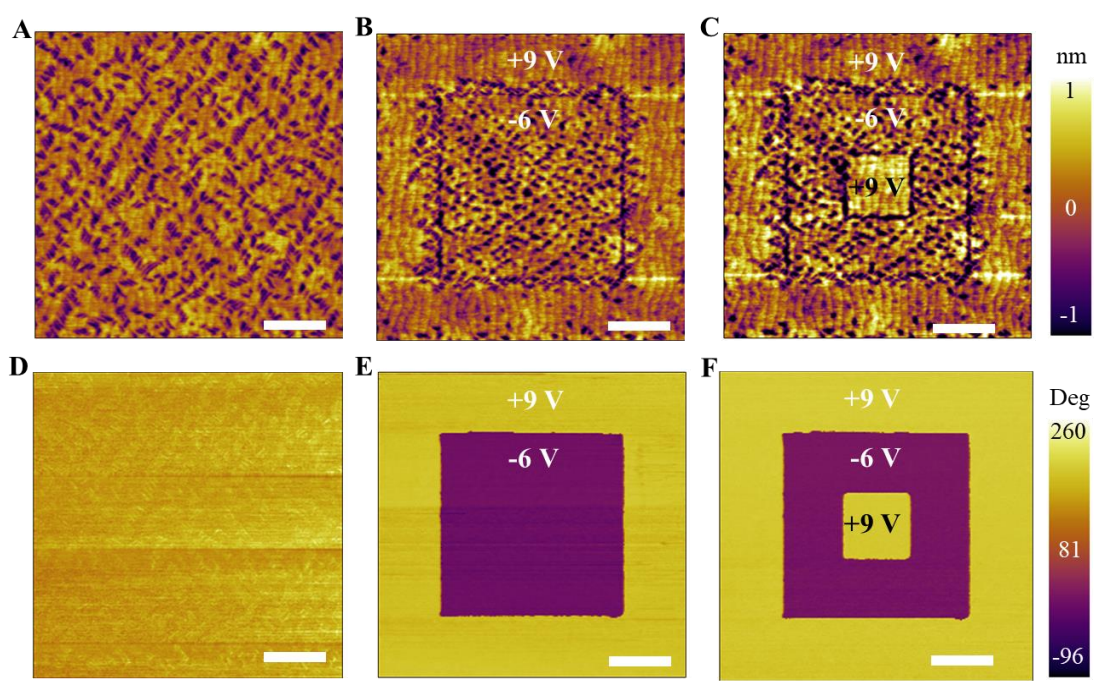

**Figure 3. Electrical Control of Phase Transition and Polarization Switching in T/R Mixed Phase BFO on SrTiO$_3$ Substrate**
(A-C) AFM images of 50 nm-thick BFO with 27 nm-thick CCMO on SrTiO$_3$ substrate demonstrate the reversible R-T phase transition under electric field.
(D-F) Out-of-plane PFM phase images corresponding to (A-C) show electrical switching of ferroelectric polarization. The scale bars in (A-F) are 1μm.

The interface CCMO layer not only provides the ability to obtain desired strains but can also serve as a bottom electrode that enables the study of the ferroelectric switching behavior of BFO under electric field. A 50 nm-thick BFO sample with T and R mixed phases on STO substrate is probed using AFM and piezoresponse force microscopy (PFM). AFM images (Figure 3 A-C) and the corresponding out-of-plane PFM phase images (Figure 3 D-F) demonstrate the reversible control of both the R-T phase transition and the ferroelectric polarization switching under electric field. The corresponding out-of-plane PFM amplitude images are shown in Figure S3. These results can also be produced in samples grown on LSAT and NSO substrates (Figure S4), which again confirmed the creation of T/R mixed phase and morphotropic phase boundary using this proposed route overcoming substrate limitations.

**Origin of imposing strain beyond substrate limitations**

To elucidate the origin of strain that can be imparted beyond substrate limitations, further structural studies of BFO (~ 7 nm)/CCMO (~ 27 nm)/STO heterostructures have been performed using X-ray reciprocal space maps (RSMs). RSM (002) reflections (Figure 4A) show the epitaxial growth of BFO and CCMO multilayers on STO substrate and demonstrate the formation of T phase and the absence of R phase in ~ 7 nm-thick BFO, which are consistent with our XRD data (Figure 1D) and AFM measurements (Figure 2A). Diffraction peaks of RSM (103) (Figure 4B) reveal the full strain relaxation of CCMO layer on STO substrate, suggesting that the origin of the desired strain to stabilize the metastable T phase beyond substrate limitations is caused by coherent growth of BFO on relaxed CCMO interface layer. The lattice parameters extracted from RSMs are a = 3.77 Å, c = 3.72 Å for CCMO and a = 3.77 Å, c =4.63 Å (c/a = 1.23) for BFO, further confirming that the relaxed CCMO layer can provide a compressive strain as much as ~ 4.8% (the in-plane lattice parameter a = 3.96 Å of ground state R-BFO) to induce R-T phase transition.

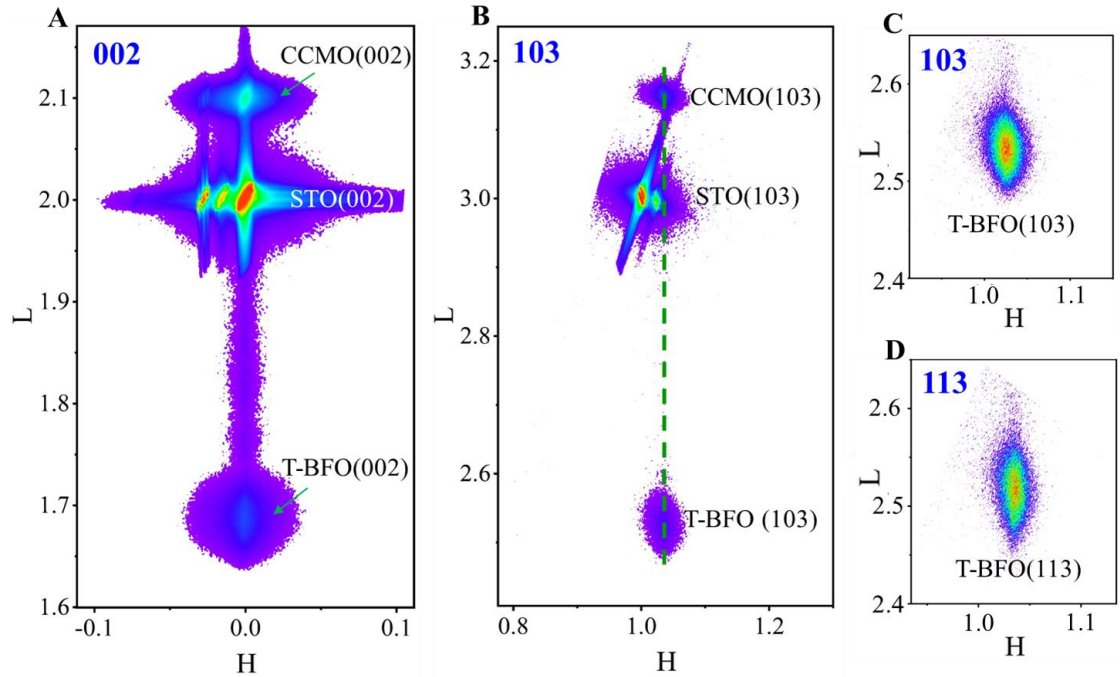

**Figure 4. Origin of Imposing Strain beyond Substrate Limitations**
(A) RSM (002) reflections of 7 nm-thick BFO grown on STO with the insertion of CCMO demonstrate the epitaxial growth of BFO and CCMO multilayers and the formation of T phase.
(B) RSM (103) reflections reveal the full strain relaxation of CCMO layer, leading to generate the desired strain to stabilize the metastable T phase beyond substrate limitations.
(C-D) Fine scans of RSM (103) and (113) reflections show single peak feature, indicating the creation of real T phase BFO.

Moreover, the BFO (103) and (113) RSM reflections (Figure 4C and 4D) reveals the realization of real T ($T_R$) phase BFO *via* this proposed method. Although other approaches such as biaxial strain induced on LAO[64] and YSZ[65] substrates, chemical doping[50] and uniaxial strain[39,40] have also been reported to obtain T-BFO, here we achieve $T_R$ phase BFO beyond substrate limitations. RSM data of 50 nm-thick BFO on STO, NSO and LSAT (as shown in Figure S5, Figure S6 and Figure S7, respectively) further demonstrate the coexistence of R and T phases in agreement with AFM measurements (Figure 2C, F and Figure S2C) and illustrate the formation of highly distorted tetragonal-like $M_C$ phase (T) phase in 50 nm-thick BFO film which exhibits three-fold splits along (103) and two-fold splits along (113). This thickness-dependent $T_R$ to T structure evolution is consistent with and can be explained by previous study[66]. The (103) RSM scans (Figure S5B, Figure S6B and Figure S7B) of BFO/CCMO heterostructures on STO, NSO and LSAT substrates also present the full strain relaxation of CCMO, giving rise to generate large compressive strain in BFO to obtain T phase.

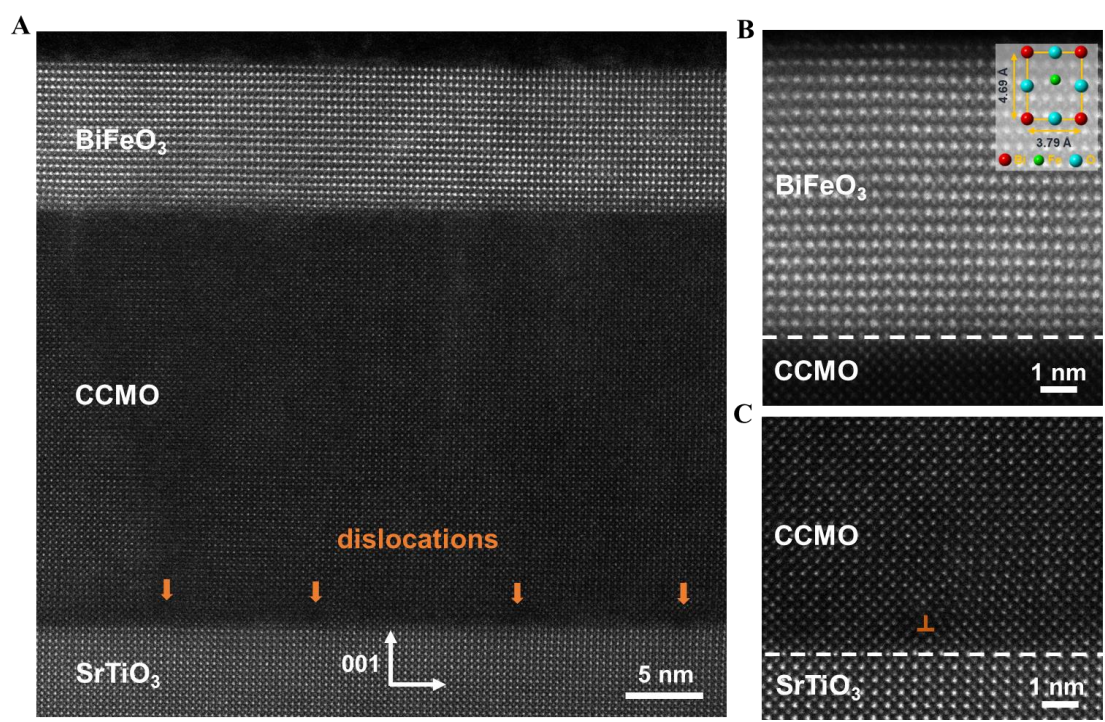

**Figure 5. Atomic Structure of BFO/CCMO Heterostructures**
(A) Dark-field STEM image of BFO (~ 7 nm)/CCMO (~ 27 nm)/STO, suggesting the strain relaxation of CCMO layer through the introduction of dislocations.
(B) Atomic scale ADF-STEM image reveals coherent growth and epitaxial stabilization of metastable true T phase BFO (c/a ~ 1.24) on STO substrate and shows abrupt interface between BFO and CCMO.
(C) High-resolution STEM image indicates the sharp interface of CCMO/STO.

To gain further understanding of the atomic structure of these strained heterostructures, we performed annular dark-field scanning transmission electron microscopy (ADF-STEM) studies on the BFO (~ 7 nm)/CCMO (~ 27 nm)/STO sample. The dark field cross-sectional STEM image (Figure 5A) demonstrates the high quality, epitaxial growth of BFO/CCMO heterostructures on STO substrate. The CCMO layer undergoes strain relaxation through the introduction of dislocations (marked with orange arrows), resulting in its bulk state with in-plane lattice parameter of a ~ 3.77 Å (calibrated by STEM image). This further confirms the mechanism of imposing a large compressive strain of -4.8% through the fully relaxed CCMO layer for the coherent growth and epitaxial stabilization of metastable $T_R$ phase BFO (c/a ~ 1.24) on STO substrate as demonstrated in the atomic scale ADF-STEM image (Figure 5B). High-resolution STEM images also reveal the abrupt interfaces of BFO/CCMO (Figure 5B) and CCMO/STO (Figure 5C). The interfacial CCMO/BFO is $Ca_{0.96}Ce_{0.04}O$ termination (Figure S8A), leading to the downward polarization orientation in BFO (Figure S8B and Figure 3D-F), in consistent with previous study[67].

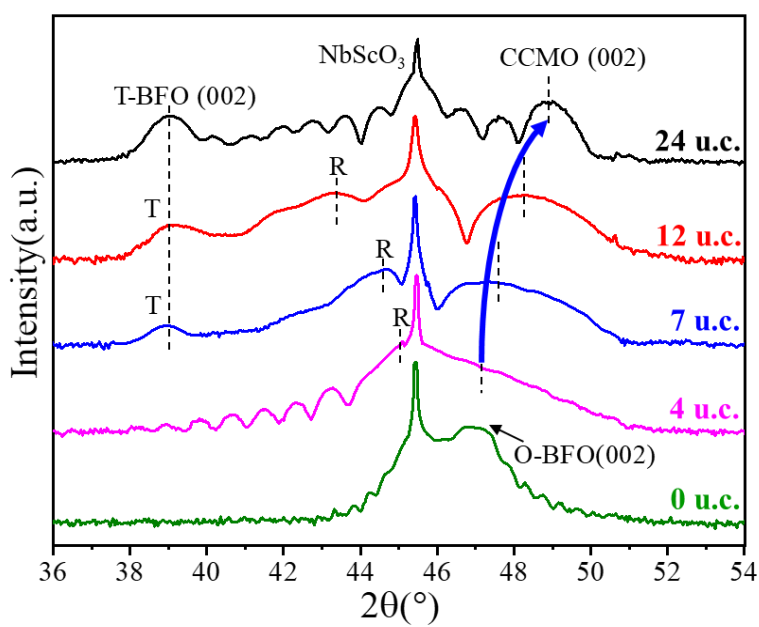

**Figure 6. Continuous Strain Tunability**
Typical x-ray θ–2θ scans of BFO/CCMO heterostructures with identical thickness (~ 14 nm) of BFO and various thickness of CCMO (0-24 unit-cells, u.c.), demonstrating that epitaxial strain can be continuously tunable by adjusting the thickness of interface CCMO layer that
14

enables the observation of strain-driven continuous O-R'-T' phase transition in BFO on a single, tensile strain NdScO₃ substrate (the fine scan shown in the insets points to the strain induced O-R' phase transition in BFO by inserting only 4 u.c. thick CCMO).

**Continuous strain tunability**

From the combination of terraced morphology (Figure 2A), RSM studies (Figure 4) and STEM results (Figure 5), we conclude that large epitaxial strain can be imparted to induce phase transition and stabilize super-tetragonal BFO phase beyond substrate limitations. Next, we turn to study the possibility of continuously tuning the epitaxial strain. Single layer 48 u.c.-thick CCMO was grown on NSO by RHEED (Reflection High Energy Electron Diffraction)-assisted PLD. The in-situ growth RHEED diffraction patterns of the CCMO film, with thicknesses of 0 u.c. ~ 48 u.c., have been captured during the growth process (Figure S9), revealing the full strain state 2 u.c.-thick CCMO film, partial strain relaxation of 4 u.c. ~ 12 u.c-thick CCMO and full strain relaxation of 24 u.c. and 48 u.c.-thick CCMO films. These results show that the strain state of CCMO on the substrate can be continuously tuned by precise control of the film thicknesses, i.e., the lattice constants of CCMO can be tuned from 3.77 Å to 4.01 Å through gradual misfit strain relaxation between CCMO and the substrate.

Consequently, by tuning the thickness of CCMO layer, tunable strain states can be imparted into the overlying BFO layer. Thus, a series of BFO/CCMO heterostructures, with identical thickness (~ 14 nm) of BFO and different thickness of CCMO (0-24 unit-cells, u.c.), were synthesized using PLD on NSO substrates. Typical XRD θ–2θ scans of these samples are displayed in Figure 6. The peak shift of CCMO with increasing thickness from 4 u.c. to 24 u.c., again indicates the gradual strain relaxation of the interface CCMO layer. This allows us to continuously tune the strain state, which is demonstrated through the strain-driven continuous O-R-T phase transition in BFO by changing CCMO thickness in Figure 6. The very narrow rocking curves (Figure S10) indicate the high structural quality of all these BFO thin films on NSO substrates with different thickness of CCMO buffer layers.

The orthorhombic (O) structure of the BFO/NSO sample (without CCMO layer) is revealed by RSM measurements (Figure S11) and further confirmed by the PFM data exhibiting in plane ferroelectric polarization (Figure S12), in consistent with previous study[58]. We also note that other reports[8,68] show different symmetries (such as monoclinic $M_B$) of BFO on large tensile substrates, which may differ depending on the precise growth conditions. Comparing to the BFO sample without CCMO (0 u.c. thick), the insertion of only 4 u.c.-thick CCMO enables the generation of sufficient compressive epitaxial strain to drive O-R phase transition of BFO. Furthermore, considerable strain can be obtained to induce the metastable T-BFO by introducing 7 u.c.-thick CCMO, resulting in the coexistence of T and R phases. With further increase in CCMO thickness from 12 u.c. to 24 u.c., the fraction of T-BFO increases while that of R-BFO decreases. As shown in Figure 1d, thicker (72 u.c. ~ 27nm) CCMO on NSO undergoes full strain relaxation through the introduction of dislocations, leading to the formation of pure T phase BFO. The typical PFM data (Figure S13) also point to the O-R-T phase transition in BFO grown on different thickness of the CCMO layer. Therefore, here we demonstrate a route to continuously tune epitaxial strain by precisely controlling the thickness of the interface CCMO layer. This allows the first observation of continuous O-R-T phase transition in BFO on a single substrate. We would like to mention that the study of the functionalities of these films might be limited as the thickness of the bottom electrode CCMO layer is different. We suggest using thinner CCMO (1-2 u.c) or a different bottom electrode for the orthorhombic BFO.

## Conclusion

We have revealed the ability to achieve desired epitaxial strain in BFO beyond substrate limitations through strain relaxation of an interface CCMO layer, enabling the stability of metastable T phase BFO not only on moderate compressive strain $SrTiO_3$ and LSAT substrates, but also on large tensile strain $NdScO_3$ substrate. Moreover, tailoring the gradual strain relaxation of CCMO, by the precise control of its thickness at nanoscale, leads to the creation of continuously tunable epitaxial strain that enables the first observation of continuous O-R-T phase transition and designing multiple morphotropic phase boundaries in BFO on a single, tensile strain $NbScO_3$ substrate. This work suggests a pathway to generate continuously tuning epitaxial strains in oxide heterostructures beyond substrate limitations and provides a platform to manipulate related ferroelectric, ferromagnetic and superconductive properties.

## EXPERIMENTAL PROCEDURES

### Film Growth

A commercial 10% Bi excess BFO target was used to compensate the Bi volatilization during the PLD growth process. The commercial CCMO target with the composition of $Ca_{0.96}Ce_{0.04}MnO_3$ was used to grow the CCMO layer. A series of BFO and CCMO thin films, BFO/CCMO heterostructures were grown on single crystal (001) LAO, (001) LSAT, (001) STO and (110)$_O$ NSO substrates by pulsed laser deposition (PLD) at 690 °C under an oxygen partial pressure of 100 mTorr and cooled in a 1 atm oxygen atmosphere. The laser fluence and repetition rate were 1.5 J/cm$^2$ and 5 Hz, respectively.

**AFM and XRD Measurements**

The AFM and PFM measurements were carried out by Pt-coated conductive probe under contact mode using an atomic force microscope (AFM)-based setup Scan Probe Microscope. The 2theta-omega and rocking curve scans were performed by the X-ray beam in 1W1A station of Beijing Synchrotron Radiation Facility with the wavelength of ~1.54848Å and the beam size of "1mm×0.6mm". Reciprocal space mappings (RSMs) were conducted using X-ray diffractometers (XRD, X'Pert Pro M, PANalytical Inc.) with wavelengths of $CuK_{\alpha1}$=~1.5406Å and $CuK_{\alpha2}$=~1.5444Å, in which the intensity ratio of $K_{\alpha1}:K_{\alpha2}$ is 2:1. The beam we selected is a parallel beam with irradiated height ~0.5mm and width ~8mm.

**TEM Characterization**

Focused ion beam lift-out techniques were performed using FEI Helios G4 UX to fabricate BFO/CCMO/STO cross-section membranes of around hundred-nanometer thickness for maintaining the same strain condition as in bulk heterostructures. Then JEOL JEM ARM 200CF with the CEOS fifth-order corrector working at 200 kV was used to image the atomic configurations in these cross-section membranes. An annular detector was set up to collect high-angle scattered electrons when a sub-angstrom electron probe scans across the sample surface, forming high-resolution "Z-contrast" images.

## SUPPLEMENTAL INFORMATION

Supplemental Information can be found online at

## ACKNOWLEDGEMENTS

We sincerely thank professor Zuhuang Chen, Dr. Chuanwei Huang, Dr. Zedong Xu, Ms. Sixia Hu and Mr. Cai Jin for the fruitful discussions. This work was supported by the National Key Research and Development Program of China (No. 2016YFA0201002) and the National Natural Science Foundation of China (Grant Nos. 11704130, U1832104 and 91963102). Authors also acknowledge the financial support of the Natural Science Foundation of Guangdong Province (Grant No. 2017A30310169) and Guangdong Science and Technology Project-International Cooperation (Grant No. 2019A050510036). D.C. thanks the financial support from the Guangzhou Science and Technology Project (Grant No. 201906010016) and Guangdong Provincial Key Laboratory of Optical Information Materials and Technology (No. 2017B030301007). Y. Z. thanks the financial support from the Research Grants Council of Hong Kong (Project No. 15305718) and the Hong Kong Polytechnic University grant (Project No. 1-ZE6G). N. W. thanks the financial support from the Research Grants Council of Hong Kong (Project No. C6021-14E). L.C. acknowledges the Science and Technology Research Items of Shenzhen (Nos. JCYJ20170412153325679 and JCYJ20180504165650580). We also acknowledge the support from 1W1A station of Beijing Synchrotron Radiation Facility and beamline BL14B at Shanghai Synchrotron Radiation Facility.

## AUTHOR CONTRIBUTIONS

D.C. conceived the project and designed the experiments. X.D and X.Y. fabricated the samples. X.C., C.X., N.W. and Y.Z. carried out the STEM measurements. C.C., X.Y., S.H. and G.T. performed the XRD measurements. Y.L., H.X., C.C., Y.C. and Z.L. contributed to the HRXRD and RSM measurements at Beijing Synchrotron Radiation Facility and Shanghai Synchrotron Radiation Facility, respectively. X.D. and F.S. carried out the PFM measurements. Z.F., M.Q., X.L. and G.Z. analyzed the AFM data. X.G., L.C. and J.-M.L. discussed the TEM data. D.C. wrote the paper with contributions and feedback from all authors. All authors discussed the results and commented on the manuscript.

## DECLARATION OF INTERESTS

The authors declare no competing interests.

Supplementary Information

# Strain engineering of epitaxial oxide heterostructures beyond substrate limitations

Xiong Deng[1,10], Chao Chen[1,10], Deyang Chen[1,2,11,*], Xiangbin Cai[3], Xiaozhe Yin[1], Chao Xu[4], Fei Sun[1], Caiwen Li[1], Yan Li[5], Han Xu[6], Mao Ye[7], Guo Tian[1], Zhen Fan[1], Zhipeng Hou[1], Minghui Qin[1], Yu Chen[5], Zhenlin Luo[6], Xubing Lu[1], Guofu Zhou[2,8], Lang Chen[7], Ning Wang[3], Ye Zhu[4], Xingsen Gao[1], Jun-Ming Liu[1,9]

[1]Institute for Advanced Materials, South China Academy of Advanced Optoelectronics, South China Normal University, Guangzhou 510006, China

[2]Guangdong Provincial Key Laboratory of Optical Information Materials and Technology, South China Academy of Advanced Optoelectronics, South China Normal University, Guangzhou 510006, China

[3]Department of Physics and Center for Quantum Materials, The Hong Kong University of Science and Technology, Clear Water Bay, Kowloon, Hong Kong, China

[4]Department of Applied Physics, The Hong Kong Polytechnic University, Hung Hom, Kowloon, Hong Kong, China

[5]Institute of High Energy Physics, Chinese Academy of Sciences, Beijing 100049, China

[6]National Synchrotron Radiation Laboratory, University of Science and Technology of China, Hefei, Anhui 230026, China

[7]Department of Physics, Southern University of Science and Technology, Nanshan District, Shenzhen, Guangdong 518055, China

[8]National Center for International Research on Green Optoelectronics, South China Normal University, Guangzhou 510006, China

[9]Laboratory of Solid State Microstructures and Innovation Center of Advanced Microstructures, Nanjing University, Nanjing 210093, China

[10]These authors contributed equally

[11]Lead Contact

*Correspondence:
deyangchen@m.scnu.edu.cn (D.C.)

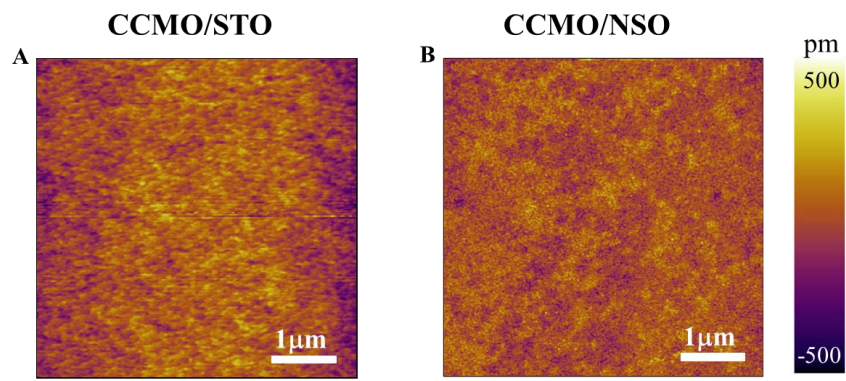

**Figure S1. High quality CCMO thin films**

AFM images of the morphology of 27 nm-thick CCMO thin film grown on STO (A) and NSO (B) substrates, showing high quality growth of flat CCMO films with a root mean square (RMS) roughness of ~ 0.2 nm.

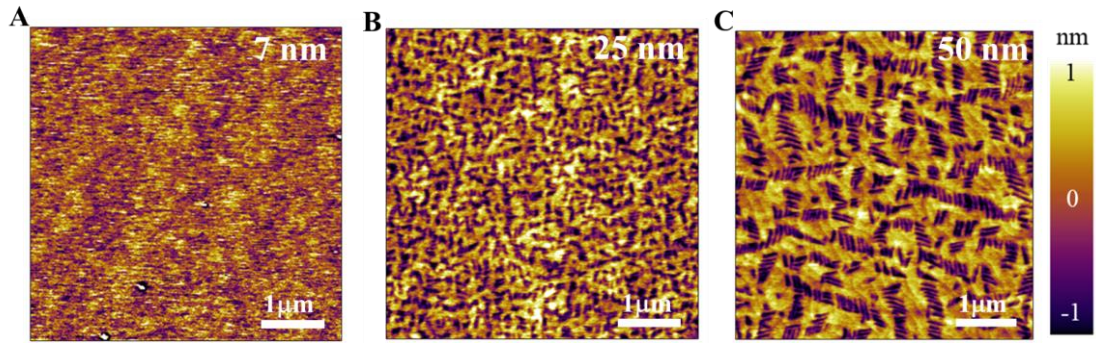

**Figure S2. Thickness-dependent phase transition on LSAT substrate**

AFM images of 7 nm (A), 25 nm (B) and 50 nm (C) thick BFO with 27 nm-thick CCMO buffer layer on LSAT substrates, indicating high quality epitaxial growth of BFO films with atomically flat morphologies (RMS roughness < 0.3 nm). Uniform contrast morphology (A) suggests the formation of pure T phase in 7 nm-thick BFO, whereas bright contrast matrix and dark contrast stripes in 25 nm-thick film (B) reveal the emergence of the R phase that coexists with the T phase due to relaxation of the epitaxial strain with increased thickness. Further strain relaxation in thicker film (50 nm) results in further increase of fraction of R phase (C).

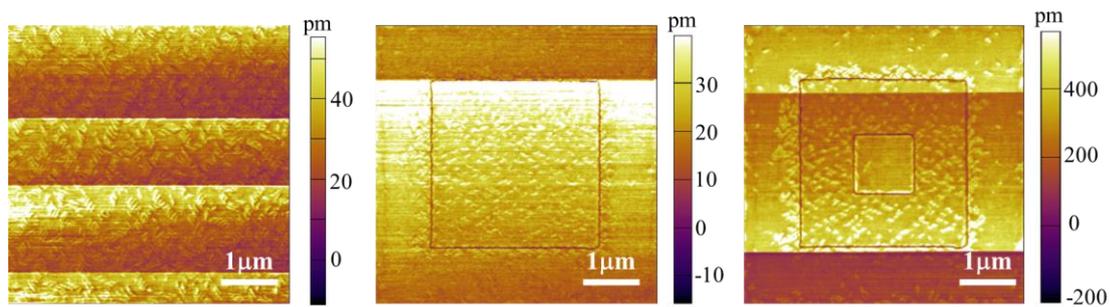

**Figure S3. Out-of-plane PFM amplitude images corresponding to Figure 3**

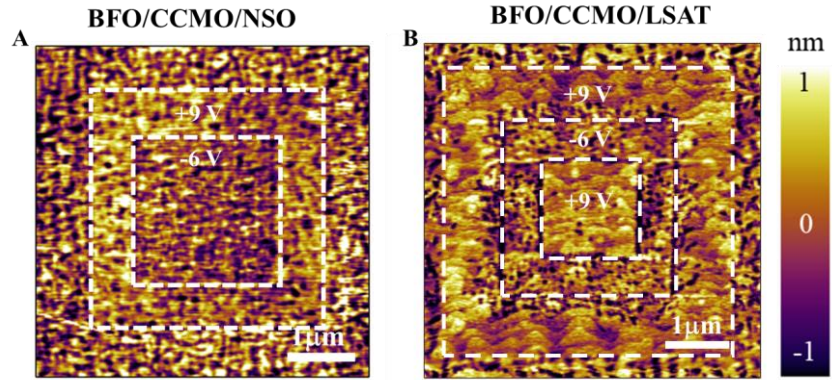

**Figure S4. Electrical control of reversible phase transition**

AFM images of 50 nm-thick BFO with 27 nm-thick CCMO bottom electrode on NSO (A) and LSAT (B) substrates show the mixed R and T phases and demonstrate the reversible R-T phase transition by applying voltage.

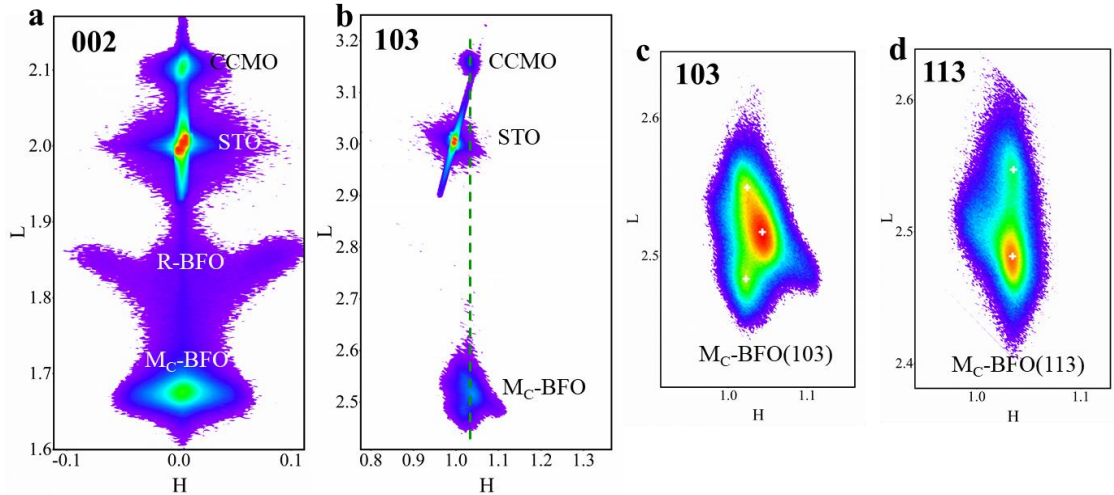

**Figure S5. Origin of imposing strain beyond substrate limitations on STO substrate**

(A) RSM (002) reflections of 50 nm-thick BFO grown on STO with the insertion of CCMO demonstrate the epitaxial growth of BFO and CCMO multilayers and the formation of T ($M_C$) and R mixed phases.

(B) RSM (103) reflections reveal the full strain relaxation of CCMO layer, leading to generate the desired strain to induce T phase and obtain T and R mixed phases on STO.

(C-D) Fine scans of RSM reflections exhibit three-fold splits along (103) and two-fold splits along (113), indicating the formation of T ($M_C$) phase BFO.

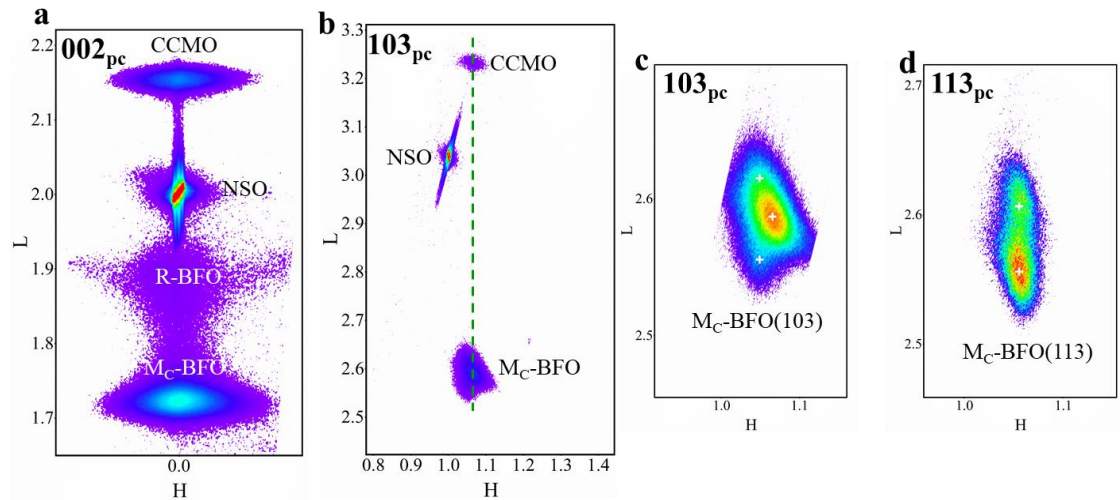

**Figure S6. Origin of imposing strain beyond substrate limitations on NSO substrate**

(A) RSM (002) reflections of 50 nm-thick BFO grown on NSO substrate with the insertion of CCMO demonstrate the epitaxial growth of BFO and CCMO multilayers and the formation of T and R mixed phases.

(B) RSM (103) reflections reveal the full strain relaxation of CCMO layer, leading to generate the desired strain to induce T phase and obtain T and R mixed phases on NSO.

(C-D) Fine scans of RSM reflections exhibit three-fold splits along (103) and two-fold splits along (113), indicating the formation of T ($M_C$) phase BFO.

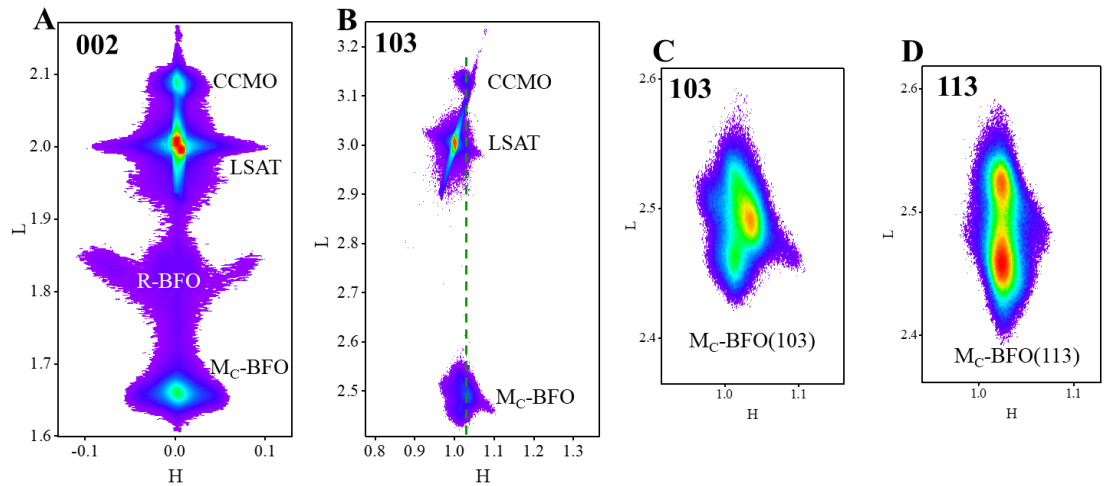

**Figure S7. Origin of imposing strain beyond substrate limitations on LSAT substrate**

(A) RSM (002) reflections of 50 nm-thick BFO grown on LSAT with the insertion of CCMO demonstrate the epitaxial growth of BFO and CCMO multilayers and the formation of T and R mixed phases.

(B) RSM (103) reflections reveal the full strain relaxation of CCMO layer, resulting in the creation of a desired strain to induce T phase and obtain T and R mixed phases on LSAT.

(C-D) Fine scans of RSM reflections exhibit three-fold splits along (103) and two-fold splits along (113), indicating the formation of T ($M_C$) phase BFO.

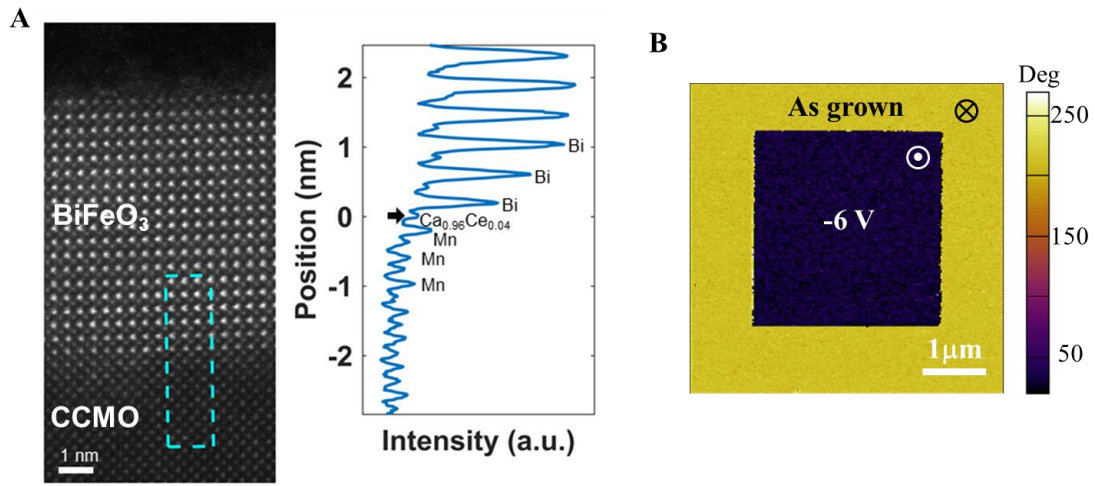

**Figure S8. Interfacial termination of BFO/CCMO**

(A) Dark-field STEM image of BFO/CCMO interface and corresponding chemical profiles across the interface.

(B) Out-of-plane PFM phase switching under -6 V voltage.

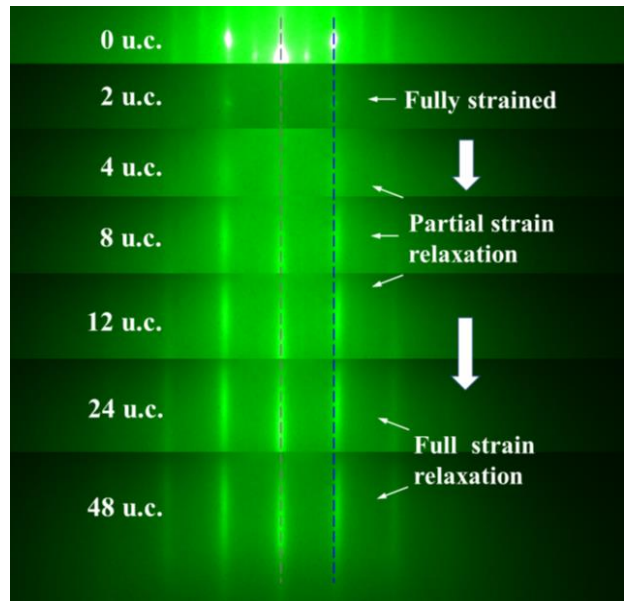

**Figure S9. In-situ RHEED diffraction patterns of 48 u.c.-thick CCMO single layer on NdScO₃ (110)$_O$ substrate**

The blue dashed line shows consistent diffraction patterns of 2 u.c.-thick CCMO film and the NSO substrate, gradual offset in 4 u.c. ~ 12 u.c-thick CCMO films, and fixed offset in 24 u.c.- and 48 u.c.-thick CCMO films, indicating fully strained state 2 u.c.-thick CCMO, partial strain relaxation in 4 u.c.- ~ 12 u.c.-thick CCMO and full strain relaxation in 24 u.c.- and 48 u.c.-thick CCMO, respectively.

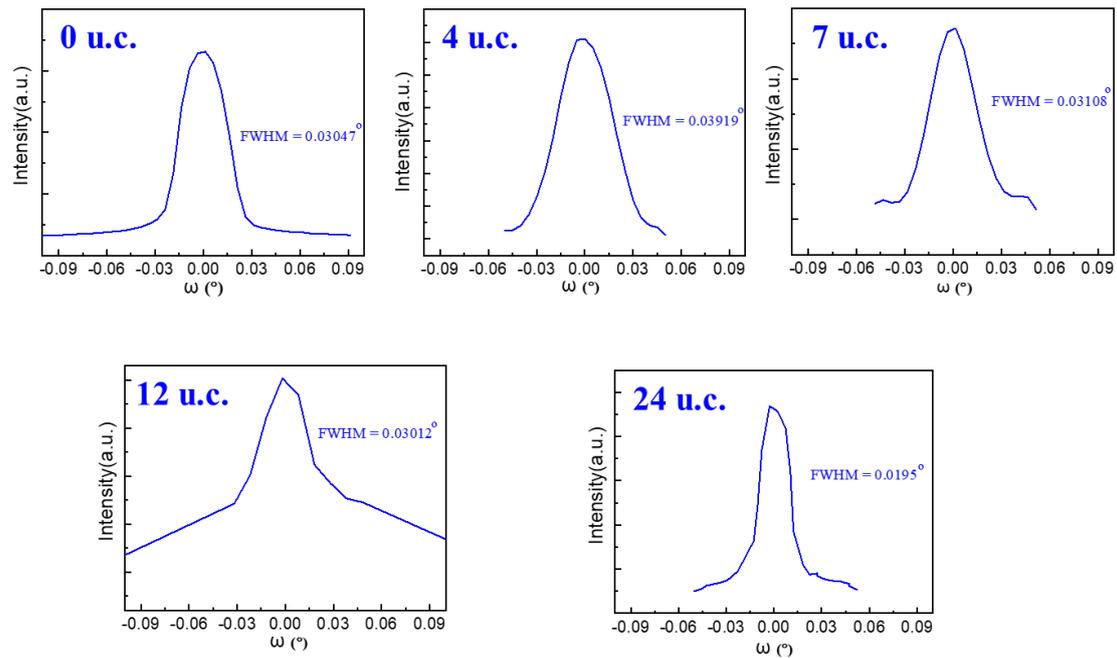

**Figure S10. Rocking curves**

Rocking curves and full width at half maximum (FWHM) of 14 nm-thick BFO grown on NSO substrates with different thickness of CCMO (0-24 u.c.). The rocking curves were detected around the 002 diffractions of BFO.

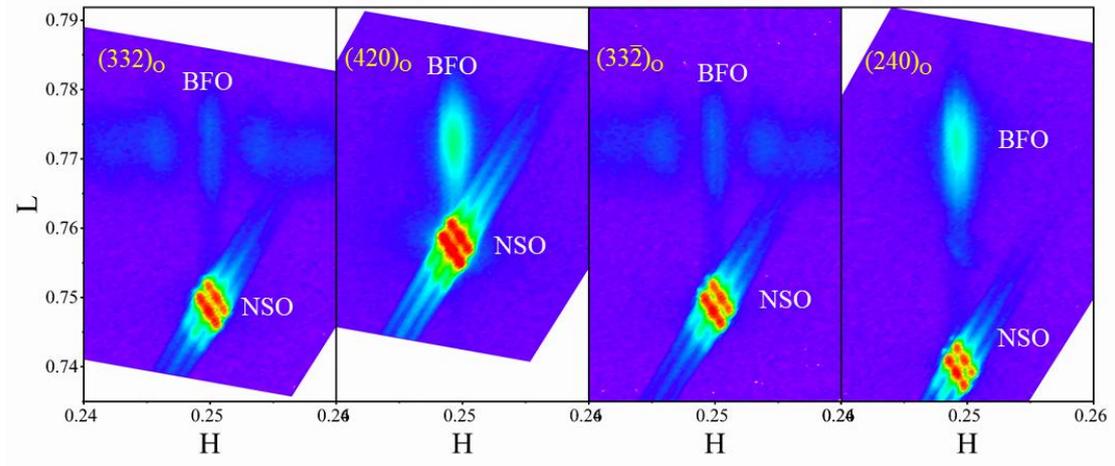

**Figure S11. Orthorhombic BFO on NSO substrate without CCMO**

The RSM data reveal the orthorhombic structure of BFO on NSO substrate without CCMO. The lattice parameters are calculated to be a=4.00 Å, b=4.01 Å, c=3.89 Å and α = β = γ = 90º.

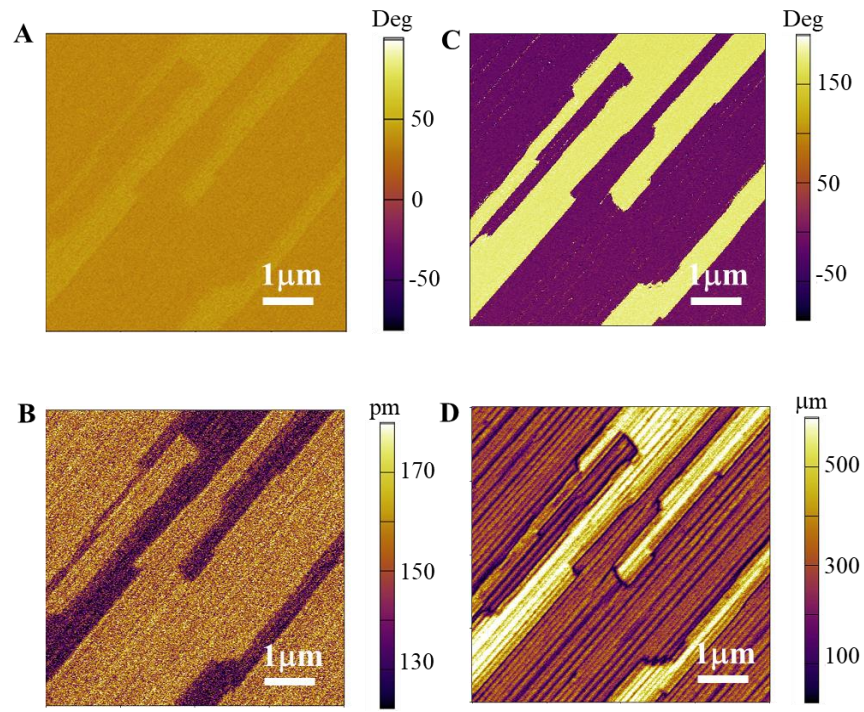

**Figure S12. PFM images of 14 nm-thick BFO on NSO substrate without CCMO**

(A) Out-of-plane PFM phase image.

(B) Out-of-plane PFM amplitude image.

(C) In-plane PFM phase image.

(D) In-plane PFM amplitude image.

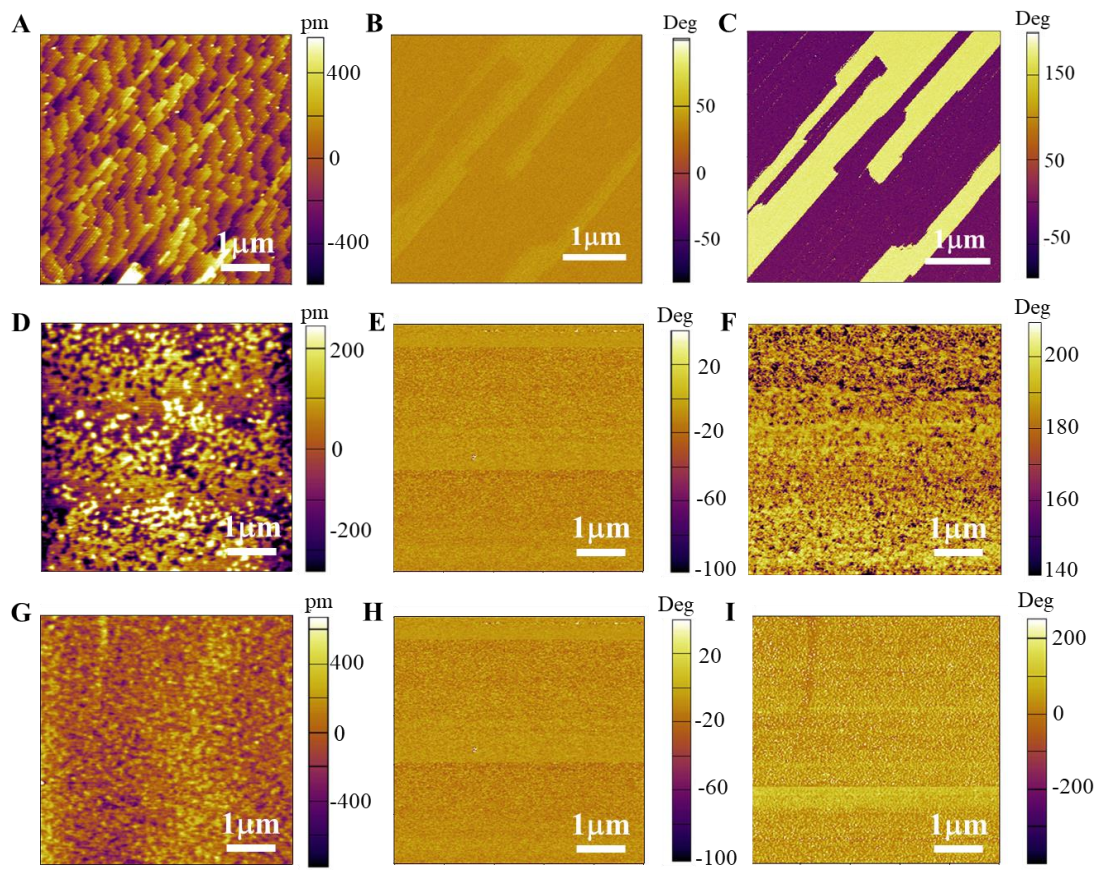

**Figure S13. Topography, out-of-plane and in-plane PFM phase images of 14 nm-thick BFO on various thickness of CCMO**

(A-C) BFO on NSO without CCMO shows uniform OOP contrast (B) while stripe domain structures in the IP image (C), suggesting the ferroelectric polarization is in the plane of the film.

(D-F) BFO on NSO with 4 u.c.-thick CCMO shows uniform OOP contrast (E) and nanodomain IP contrasts (F), pointing to the formation of R phase.

(H-J) BFO on NSO with 24 u.c.-thick CCMO shows uniform OOP (I) and IP PFM contrasts (J), suggesting the formation of T phase with single domain state.